\documentclass[12pt]{article}
\usepackage{amsfonts,amsmath,amssymb,epsf,hyperref,cite,ulem}
\usepackage[lmargin=2.0cm, rmargin=1.5cm,tmargin=2.50cm,bmargin=2.50cm]{
geometry}
\usepackage{graphicx,color,subfigure}
\usepackage[usenames,dvipsnames]{pstricks}
\usepackage{mathrsfs}
\catcode `@=11 \@addtoreset{equation}{section} \catcode `@=12

\def\vx{{\vec{x}}}

\def\vk{{\vec{k}}}

 \def\CI{{\cal I}}
\def\cD{{\cal D}}

\def\nn{\nonumber}

 \def\ep{\epsilon}

 \def\ep{{\epsilon}}

 \def\b{{\beta}}

 \def\CI{{\cal I}}
 \def\cO{{\cal O}}

 \def\ben{\begin{equation}}
\def\een{\end{equation}}
\def\beq{\begin{eqnarray}}
\def\eeq{\end{eqnarray}}
\def\bea{\begin{eqnarray}}
\def\eea{\end{eqnarray}}
\def\vx{{\vec{x}}}

\def\lds{L_{dS}}
\def\rat{\left( \frac{\lds}{-T} \right)}
\def\inrat{\left( \frac{-T}{\lds} \right)}
\def\tphi{{\tilde{\phi}}}
\def\lak{\rho}

\def\CI{\mathscr{I}}
\def\Cz{\mathcal{Z}}
\def\eq#1{(\ref{#1})}
\def\lan{\langle}
\def\ran{\rangle}

\begin{document}
\begin{titlepage}
\thispagestyle{empty}
\begin{flushright}
UK/13-12\\TIFR/TH/13-18
\end{flushright}

\bigskip

\begin{center}
\noindent{\Large \textbf
{Double Trace Flows and Holographic RG in dS/CFT correspondence}}\\
\vspace{1.0cm} \noindent{
Diptarka Das $^{(a)}$\footnote{e-mail: diptarka.das@uky.edu},
Sumit R. Das $^{(a)}$\footnote{e-mail: das@pa.uky.edu} and
Gautam Mandal $^{(b)}$\footnote{e-mail: mandal@theory.tifr.res.in}

\vspace{0.8 cm}
 $^{(a)}$ {\it Department of Physics and Astronomy, \\
 University of Kentucky, Lexington, KY 40506, USA}\\
$^{(b)}${\it Department of Theoretical Physics,\\
Tata Institute of Fundamental Research,\\
Homi Bhabha Road, Mumbai 400005, INDIA}}
\end{center}

\vspace{0.3cm}
\begin{abstract}

If there is a dS/CFT correspondence, time evolution in the bulk
should translate to RG flows in the dual euclidean field theory. 
Consequently, although the dual field is expected to be non-unitary,
its RG flows will carry an imprint of 
the unitary time evolution in the bulk. In this note we
examine the prediction of holographic RG in de Sitter space for the
flow of double and triple trace couplings in {\em any} proposed dual.
We show quite generally that the correct form of the field theory beta functions
 for the double trace couplings is obtained from holography, provided one identifies 
the scale of the field theory with $(i|T|)$ where $T$ is the `time' in conformal coordinates. For $dS_4$, we find that with an 
appropriate choice of operator normalization, it is possible
to have real $n$-point correlation functions as well as beta functions with real coefficients.  This choice leads to an RG flow with an IR fixed point at {\em negative} coupling unlike in
a unitary theory where the IR fixed point is at positive coupling.  The proposed correspondence of $Sp(N)$
vector models with de Sitter Vasiliev gravity provides a specific example
of such a phenomenon. For $dS_{d+1}$ with even $d$, however, we find that
no choice of operator normalization exists 
which ensures reality of coefficients of the beta-functions
as well as absence of $n$-dependent phases for various $n$-point 
functions, as long as one assumes real coupling constants in the bulk Lagrangian.

\end{abstract}
\end{titlepage}
\newpage

\tableofcontents

\section{Introduction}

The dS/CFT correspondence \cite{Witten:2001kn, Strominger:2001pn,
  Maldacena:2002vr} proposes that quantum gravity in asymptotically de
Sitter space is dual to a {\em Euclidean} conformal field theory which
lives on $\CI^+$ or $\CI^-$. Specifically, it has been proposed that
the partition function of the CFT deformed by single trace operators
(which equals the generating functional for correlators of the CFT) is
the Bunch-Davies wavefunctional obtained by performing the bulk path
integral with Dirichlet boundary conditions on $\CI^+$ and Bunch-Davies condition in the infinite past. Unlike
in AdS/CFT \cite{Maldacena}-\cite{AdSR}, the meaning of this
correspondence is not completely clear, particularly because of the
difficulty in defining observables in de Sitter space \cite{Witten:2001kn}. 
While these issues are obviously important, one can nevertheless
perform computation in the dS bulk where gravity is treated semiclassically
\cite{Maldacena:2002vr}.
Keeping this in view, in this 
note we will address the question: if a dS/CFT correspondence does exist, what does it say about the dual field theory? 

To begin with, the dual field theory cannot be unitary in the usual sense \cite{Maldacena:2002vr,strom1}. The symmetry group
of the putative $d$-dimensional Euclidean CFT, $SO(d+1,1)$, is the isometry group
of both $dS_{d+1}$ and Euclidean $AdS_{d+1}$. If the CFT is unitary,
one would expect that the dual is a bulk theory living in Euclidean
$AdS_{d+1}$. Thus, the CFT dual to $dS_{d+1}$ is non-unitary.
On the other hand, there is a unitary time evolution in the
$dS_{d+1}$ bulk (examples of which 
we will consider explicitly below); if the holographic correspondence
is true, this will clearly imply some constraints on the dual
field theory. In this note, we will explore these constraints on the
RG flow of double and triple trace deformations in
the dual field theory.  For double trace couplings, the story for AdS
is well known \cite{Witten:2001ua, Mueck:2002gm, Vecchi:2010dd}: for a
relevant deformation with positive coupling, the theory flows into a
IR fixed point, in complete agreement with the prediction of the dual
large-N field theory.

We will calculate the beta function for the double and triple trace 
couplings of a proposed CFT dual to de Sitter space using the holographic renormalization group techniques of \cite{Heemskerk:2010hk} and
\cite{Faulkner:2010jy} (for previous  work on the subject, see
\cite{Akhmedov:1998vf}-\!\!\cite{Skenderis:2002wp}). We will show that the 
beta function has the same structure as that expected from general field theory considerations, along with  holographically determined coefficients.
 In particular the coefficient of the quadratic term of the
double trace beta function equals the normalization of the two point function; similar statements are true for the
triple trace beta function. For $dS_4$, we find that  the specific
choice of operator normalization which leads to
real $n$-point correlation functions  \cite{strom1} 
also leads to beta functions with real coefficients. 
 This leads to a beta function whose quadratic term 
differs in sign from that in Euclidean $AdS_4$, so that
the IR fixed point now appears at negative rather than positive
coupling. 
The recent proposal of a duality between $Sp(N)$ vector
models in three Euclidean dimensions and Vasiliev theory in $dS_4$
\cite{strom1}-\!\!\cite{Anninos2} provides a specific realization of the
above result.

For $dS_{d+1}$ with even $d$, however, we find, first of all, that
no choice of operator normalization exists which ensures
absolute  reality of 
the $n$-point functions; furthermore,
any choice of operator normalization which ensures reality of coefficients of the beta-functions forces us to have $n$-point functions  with very
specific $n$-dependent complex phases, 
$\lan O_1 \cdots O_{n} \ran   \sim i^{(n-2)(1-d)/2}$
as explained in  Section \ref{sec:complex}. 
These assertions are proved in Section \ref{sec:complex}
under the general condition of real coupling constants in the bulk Lagrangian.
It is important to note that the reality of the coefficients
of the bulk Lagrangian, which is tied to the unitarity of the bulk field theory,  
plays a crucial role here.

\section{The main result}

In this section we first derive the field theory beta function at leading order of $1/N$. We then summarize our findings for the holographic beta function.

\subsection{\label{sec:ft2pt}Field theory: 2-pt function vs. double trace beta-function}

Consider the two-point function of an operator 
$\cO(x)$ in a $d$-dimensional {\em Euclidean} CFT:
\ben
\langle \cO(k_1) \cO(k_2) \rangle_0 = G_0(k) (2\pi)^d \delta(k_1 + k_2),
\; G_0(k)= b k^{-2\nu}, \; 2\nu \equiv d - 2 \Delta
\label{green-0}
\een
where $\cO$ is a scalar operator of dimension
$\Delta$ \footnote{  
\label{ftnt:relevant}In the context of (A)dS/CFT, we will  
consider alternative
quantization, where $\cO$ will be identified with
$\cO_-$, as in (\ref{1-13}). In that case, $\Delta
= \Delta_-$ (see \eq{1-5}), and the value of $\nu$ follows the 
usual definition. Among other things, the choice of
alternative quantization ensures that the double trace
flow is relevant.}. 
The exponent of $k$ follows from dimensional analysis; the
subscript $0$ implies that the correlator is computed in the
unperturbed CFT. The constant $b$ denotes the normalization 
of the operator $\cO$.

In the following we will assume that, for large central
charge $c$ of the CFT, the leading contribution to the 2n-point function of 
$\cO$ has a factorized form (similar to Wick's theorem):
\ben
\langle \cO(k_1) \cO(k_2)....\cO(k_{2n} \rangle
=\left[ \sum_{\rm permutations} \langle \cO(k_{i_1}) \cO(k_{i_2}) \rangle ... 
\langle \cO(k_{i_{2N-1}}) \cO(k_{i_{2N}}) \rangle \right] + ...
\label{factorization}
\een
where the ... terms at the end denote $O(1/c)$ corrections.
Well-known CFT's with such properties are conformal large $N$ gauge theories
with $\cO$ a single trace operator  (or conformal large $N$ vector theories
with $\cO$ some appropriate bilinear of vectors)\footnote{ For more
general examples, see, e.g. \cite{ElShowk:2011ag}.}.

This has the following consequences:

\begin{enumerate}

\item{} The dimension of the ``double trace'' operator $\cO^2$ is  $2 \Delta$.
\footnote{We will call $\cO$ and $\cO^2$ ``single trace'' and  ``double
trace'' operators, respectively, by analogy with large $N$ gauge theories;
however, at least for the purposes of this section, this only implies
the factorization property (\ref{factorization}).} 

\item{} Under a double trace deformation ($f_0$ is a bare coupling) 
\ben
S= S_{0} + \frac{f_0}{2}  \int d^d x \cO(x)^2  
\label{double}
\een
the Green's function (\ref{green-0}) changes to \footnote{This is 
easy to derive by expanding $\exp[-S]=\exp[-S_0](1- S_{int} +
\frac12 S_{int}^2 - ...)$, and using (\ref{factorization}).}
\ben
G_f(k)= G_0(k)- f_0 G_0(k)^2 + ...= \frac{G_0(k)}{1 +  f_0 G_0(k)}
\label{green-f}
\een
{We will derive the same equation in \eq{2-4a} from a dS bulk dual.}

\end{enumerate}

\noindent The above Green's function implies 
the following `running coupling constant' \footnote{We define
the running coupling $f(k)$ by
$G_f(k)=: G_0(k) - f(k) G_0(k)^2$ (thus $f(k)$ represents the
Dyson Schwinger sum of an infinite number of Feynman diagrams in
the middle expression of \eq{green-f}).}
\ben
f(k) = \frac{f_0}{1+ f_0 G_0(k)}
\label{running}
\een
Let us define a dimensionless renormalized
coupling $\lambda(\mu)$ by the relation \footnote{Note that 
$f(k)$ is of dimension $2\nu \equiv d- 2\Delta $ since
$\cO^2$ is of dimension $2\Delta$.} 
\ben
f(\mu) =  \lambda(\mu) \mu^{2\nu}
\label{dim-less}
\een
By using the above  equations, we get
\[
\lambda(\mu)= \frac{f_0}{\mu^{2\nu} +  f_0 b}
\]
Since $f_0$ is a bare coupling, it should not depend on $\mu$.
By differentiating the above with respect to $\mu$, we get 
\ben
\mu \frac{d \lambda(\mu)}{d\mu} = -2\nu \lambda + 2\nu b \lambda^2 
\label{beta}
\een
At this stage the constant $b$ is arbitrary and is not
necessarily real; holography allows us to determine the value of
$b$, as  in
\eq{1-14a} (for a dS/CFT) where $b$ is complex and \eq{1-14c} 
(for AdS/CFT) where $b$ is real and positive.  
For unitary theories, on general grounds,
$b$ must be real and positive and we have the well-known result that the theory flows to a IR
fixed point at positive coupling. 

Note that we have arrived at (\ref{beta})
with minimal assumptions about the CFT and about the operator
$O$ (essentially its scaling and factorization). 

\subsection{Bulk dual}

Let us now assume that our CFT has a bulk dual.  The $SO(d+1,1)$
conformal symmetry implies that the bulk must be either $AdS_{d+1}$ or
$dS_{d+1}$.  A double trace deformation then translates to modified
boundary conditions for the dual bulk field
\cite{Witten:2001ua}. For $\nu > 0$ in (\ref{green-0}) the deformation has to be around alternative quantization. 
Following the procedure of integrating out
geometry devised in \cite{Heemskerk:2010hk} and \cite{Faulkner:2010jy}
we will derive the beta-function of the field theory from bulk
Schrodinger equations. For $AdS$ the time in the Schrodinger equation
is euclidean and identified with the radial coordinate, which is identified with the RG scale of the field theory:  this derivation is already contained in \cite{Heemskerk:2010hk,Faulkner:2010jy}. For $dS$,  bulk evolution is in real time, and the precise relationship of time with the field theory scale is less clear. If $T$ denotes the bulk time in inflationary coordinates (which in our convention is negative), we will find that the beta function (\ref{beta}) is again reproduced, {\rm provided} we identify $(-iT)$ with the RG scale of the dual theory.

We will find below that, the equation (\ref{green-f}) is
reproduced holographically both in the case of AdS and dS (see \eq{2-4}
and \eq{2-4a}). Further, with the above 
holographic identification
of the field theory cut-off, the beta-function \eq{beta} is reproduced
exactly in both cases. For $dS$, 
unlike in $AdS$ we cannot demand that $b > 0$ or even real in the field theory. However,  for $dS_4$ it was argued in \cite{strom1} that the only way to 
ensure real $n$ point functions is to have $b$ real and negative. This is the normalization used in
\cite{Maldacena:2002vr} as well.  This leads to the conclusion that the IR fixed point of the dual theory is at {\em negative} coupling. 
This is consistent with the conjecture of \cite{strom1}: indeed a calculation of the beta function of $Sp(N)$ field theory leads to the same beta function (this has been calculated to one loop in \cite{LeClair:2007iy}). 

However, for $dS_{d+1}$ with even $d$, as explained at the
end of the Introduction, reality of $b$ is only
possible if one allows for  specific
$n$-dependent complex phases of the  $n$-point correlation functions
(see Section \ref{sec:complex} for details).

\section{Holographic dictionaries}

\subsection{dS/CFT dictionary}
We will consider the inflationary patch of $dS_{d+1}$ with a metric
\ben
ds^2 = \frac{\lds^2}{T^2} \left[ - dT^2 + d\vx^2 \right]
\label{1-1}
\een
with $-\infty \leq T \leq 0$ 
We will consider a massive minimally
coupled scalar in this geometry with the action
\ben
S_\epsilon = S_{gr}+ \frac{1}{2 G_N}\int_{-\infty}^\epsilon dT \int d^{d}x \rat^{d+1} \left[
  \inrat^2 [(\partial_T\phi)^2 -(\nabla \phi)^2] - m^2 \phi^2 \right]
\label{1-2}
\een
where $S_{gr}$ is the gravity action and $\epsilon$ is a cutoff. In
the following we will consider the dynamics of the scalar - we will
therefore drop the gravity part. We will work in a probe approximation and ignore the backreaction on gravity. A bulk wavefunction can be now
defined by the path integral
\ben
\Psi[\phi_0(\vx),\epsilon] = \int_{\phi(\epsilon,\vx) = \phi_0(\vx)} \cD
\phi(T,\vx) {\rm{exp}}\left( iS_\epsilon \right)
\label{1-3}
\een
where the field satisfies Bunch-Davies conditions at $T =
-\infty$. $S_\epsilon$ is the action obtained by integrating from
$T=-\infty$ to $T=\epsilon$, and $\epsilon < 0$.

In the following we will use a notation
\ben
\rho \equiv \sqrt{\frac{L_{dS}^{d-1}}{G_N}}
\label{0-1}
\een

The dS/CFT correspondence as interpreted in \cite{Maldacena:2002vr,
  Harlow:2011ke,strom1,Anninos2}
then claims that this
wavefunctional is related to the partition function of a dual CFT in
the presence of a source. More precisely, in the standard quantization
of the CFT 
\begin{align}
& \lan{\rm exp} \left[\int d^dx\,  \phi_0 (\vx)\,
\Cz(\epsilon) \, \cO_+(\vx) \right] \ran_{st} = \Psi [\phi_0(\vx),\epsilon], ~~~
\Cz(\epsilon) =   \frac{\lak}{\sqrt{\gamma}} (-i\epsilon)^{-\Delta_-}
\label{1-4}
\end{align}
where
\ben
\Delta_\pm = d/2 \pm \nu, ~ \nu \equiv \sqrt{d^2/4 - m^2L_{dS}^2}
\label{1-5}
\een Here $\Cz(\epsilon)$ is a normalization factor used to define the GKPW
relation \eq{1-4}. The important part of this factor is the numerical
coefficient $\gamma$ which we treat {\it a priori} to be complex. This
constant is taken to be $\gamma = 1$ in
\cite{Harlow:2011ke,strom1,Anninos2}.  We will come back to a detailed
discussion of this coefficient later.  Note that the factor
$(-i\epsilon)$ is naturally identified with the field theory UV cutoff
\cite{strom1}.

We will be concerned with the semiclassical limit where the functional
integral on the right hand side of (\ref{1-3}) can be evaluated by
saddle point.  The classical solution which satisfies the Bunch-Davies
condition at $T = -\infty$ and the specified boundary condition at
$T=\epsilon$ is given, in momentum space, by \ben \phi(T,k) = \left(
\frac{T}{\epsilon} \right)^{d/2} \frac{ H_\nu^{(2)}(-k
  T)}{H_\nu^{(2)}(-k\epsilon)} \phi_0(\vk)
\label{1-9}
\een
where $\nu$ is given by \eq{1-5}.
This leads to the following on-shell action
\ben
iS_{on} = -\frac{i}{2 G_N} \int [dk]~ L_{dS}^{d-1} \bigg(
\frac{\Delta_- }{(-\epsilon)^d} - \frac{k\epsilon
  H_{\nu-1}^{(2)}(-k\epsilon)}{(-\epsilon)^d H_{\nu}^{(2)}(-k\epsilon)}
\bigg)\phi_0(\vk)\phi_0(-\vk)
\label{1-12a}
\een
At late times $k|\epsilon| \ll 1$
\bea
iS_{on} &=& -i \frac{\rho^2}{2} \int [dk] \bigg( \frac{\Delta_-}{(-\epsilon)^d} - \frac{\Gamma(1-\nu)}{\Gamma(2- \nu)} \frac{k^2}{2(-\ep)^{d-2}} \bigg)\phi_0(\vk) \phi_0(-\vk) \nn \\
&+& \frac{\rho^2}{2} \int [dk]
\phi_0(\vk)\phi_0(-\vk) (-i\epsilon)^{-2\Delta_-}~H(k)
\label{1-10}
\eea
where
\ben
H(k) =  (i)^{d-1} C_1(\nu) k^{2\nu},
~~~  C_1(\nu) \equiv - 2\nu \frac{\Gamma(1-\nu)}{\Gamma(1+\nu)}
  2^{-2\nu} 
\label{1-11}
\een
In the semiclassical limit the wavefunction is then
\ben
\Psi[\phi_0(\vx),\epsilon] \sim {\rm exp} [iS_{on}]
\label{1-10aa}
\een
It may be easily checked that at early times $k|\epsilon| \gg 1$ this
reproduces the ground state of a bunch of harmonic oscillators with
``coordinates'' $\chi_\epsilon (k) = (-\epsilon)^{\frac{1-d}{2}}\phi(k)$.
At late times $k|\epsilon| \ll 1$ we need to remove the divergent piece by holographic
renormalization and define the wavefunction by 
\ben
\Psi[\phi_0(\vx),\epsilon] \sim {\rm exp} [iS_{on}^\prime]
\label{1-10a}
\een
where
\ben
iS_{on}^\prime = \frac{\lds^{d-1}}{2G_N}\int [dk]
\phi_0(\vk)\phi_0(-\vk) (-i\epsilon)^{-2\Delta_-}~H(k)
\label{1-10b}
\een
is the finite part of the on-shell action. The divergent first term 
in (\ref{1-10}) has to be removed by addition of a
counterterm to the action. Using (\ref{1-4}) 
the two point correlator of the dual operator $\cO_+$,
is given by
\ben
\lan \cO_+(k) \cO_+(-k)\ran_{st} =  G_{st}(k) = \gamma H(k)= \gamma \,
i^{d-1} C_1(\nu)\, k^{2\nu}
\label{1-11a}
\een

We will be interested in {\em alternative} quantization. The
generating functional for correlators in the appropriate CFT in this
case is obtained by extending the corresponding prescription in AdS \cite{Klebanov:1999tb},
\ben
\lan \exp\left[ \int d^dx J(\vx) \cO_- (\vx) \right] \ran_{alt} =
\kern-5pt 
\int \cD \phi_0(\vx)  \lan{\rm{exp}} \left[ 
\int d^dx \phi_0(\vx) \, \Cz(\epsilon)\,
\cO_+ (\vx)\right]\ran_{st} 
\,  {\exp}\left[ \Cz(\epsilon) \kern-4pt \int d^dx \frac{J(\vx)}{2\nu}
   \phi_0(\vx) \right]
\label{1-13}
\een
In the semiclassical approximation we may replace the generating
functional of standard quantization by the wavefunction
\eq{1-10a}.
Performing the
$\phi_0$ integral leads to a two point correlator in alternative
quantization 
\ben
G_{alt}(k) = \frac{\delta^2}{\delta J(k) \delta
  J(-k)}\lan e^{ \int d^dx J(\vx) \cO_- (\vx) } \ran_{alt}=  -\frac{1}{(2\nu)^2 
G_{st}(k)} 
\label{1-14}
\een
This inverse relation between the Green's function is exactly the same as in AdS/CFT \cite{Klebanov:1999tb}.  
Combining (\ref{1-14}),(\ref{1-11a}) and
(\ref{1-11}) we get 
\ben
\lan \cO_-(k) \cO_-(-k) \ran_{alt} = G_{alt}(k) = \frac{i^{1-d}}\gamma~ 
C(\nu) k^{-2\nu},
\;  C(\nu) \equiv  \frac{2^{2\nu}}{(2\nu)^3 
  }\frac{\Gamma(1+\nu)}{\Gamma(1-\nu)}
\label{c-nu}
\een
Comparing with (\ref{green-0}), we get the following holographically
determined value of $b$: 
\ben
b_{dS}= \frac{ i^{1-d}}\gamma~ C(\nu)
\label{1-14a} 
\een
In case of $dS_4$, Ref. \cite{strom1} chose $\gamma=1$ in keeping with
the reality of the $n$-point functions, which was also
reproduced by a CFT calculation using $SP(N)$.  However, in this paper
we are dealing with $dS_{d+1}$ for arbitrary $d$ and will keep $\gamma$
arbitrary and in principle complex. We will come back
to the important issue of the phase of $\gamma$ (equivalently
of $\Cz$) and its relation to the phases of the $n$-point 
functions and beta-function coefficients in
detail in Section \ref{sec:complex}. 

The relationship (\ref{1-13}) can be inverted to rewrite the
Bunch-Davies wavefunction in terms of the generating functional in
alternative quantization, 
\ben 
\Psi [\phi_0(\vx), \epsilon ] = \int
\cD J(\vx)~ {\rm{exp}} \left[ -\Cz(\epsilon) \int d^dx
  \frac{J(\vx)}{2\nu} \phi_0(\vx)\right] 
 \langle {\rm{exp}} \left[ \int
  d^dx J(\vx) \cO_- (\vx) \right] \rangle_{alt}
\label{1-14b}
\een

\subsection{The formulae for $AdS$}

It will be useful to record the corresponding well known formulae in euclidean $AdS$ space. The GKPW prescription for the generating functional for correlators in standard quantization reads
\ben
\lan{\rm exp} \left[\int d^dx (\epsilon)^{-\Delta_-}  \phi_0 (\vx)
\tilde\Cz(\epsilon)  \cO_+(\vx) \right] \ran_{st} = Z [\phi_0(\vx),\epsilon]
\quad
\tilde \Cz(\epsilon) \equiv 
\frac{\lak}{\sqrt{\gamma}} (\epsilon)^{-\Delta_-} 
\label{1-4b}
\een
where we of course need to replace $L_{dS} \rightarrow L_{AdS}$.
There are no factors of $i$ in the formulae, the rescaling factor $\gamma$ has to be real, the  Hankel functions are replaced by Modified Bessel functions and the quantity in square brackets in (\ref{1-11}) is the boundary Green's function in $AdS_{d+1}$ leading to the proportionality constant
\ben
b_{AdS}= \frac1\gamma~ C(\nu)
\label{1-14c} 
\een
where $C(\nu)$ is defined in \eq{c-nu}. 
Since everything needs to be real, (\ref{1-4b}) requires $\gamma$ to be real and positive, leading to a real positive $b_{AdS}$. Finally, the analog of (\ref{1-14b}) for $AdS$ may be obtained by replacing $(-i\epsilon) \rightarrow \epsilon$.

\section{Double Trace deformations}

In the following we will be interested in the deformation of the CFT
dual to alternative quantization in $dS_{d+1}$ by a double trace
operator. {The Euclidean field theory action is given by \eq{double}.
As argued in Sec \ref{sec:ft2pt}, to leading order in large N, the dimension of $\cO^2$ is then
$2\Delta$. We require the perturbation to be relevant, which means
that the CFT action $S_0$ must correspond to alternative quantization
(see also footnote \ref{ftnt:relevant}),
ensuring that $2\Delta = 2\Delta_- < d$ (see \eq{1-5})}. The generating function for
correlators in the presence of the deformation may be now written
using a Hubbard-Stratanovich transformation,
\ben
\langle
{\rm{exp}} \left[ \int d^dx J(\vx) \cO(\vx) \right] \rangle_{alt}^{f_0} = 
\int \cD\sigma
{\rm{exp}} \left[ \frac{1}{2f_0}\int d^dx \, \sigma(\vx)^2
  \right] \langle{\rm{exp}} \left[ \int d^dx (J(\vx) +
  \sigma(\vx)) \cO(\vx) \right]\rangle_{alt}
\label{2-2}
\een
{where the notation $\langle ... \rangle_{alt}^{f_0}$
denotes correlations in presence of the double trace deformation
\eq{double}.}
Using (\ref{1-4}),(\ref{1-10a}) and (\ref{1-13}) we get
\ben
 \langle{\rm{exp}} \left[ \int d^dx J(\vx) \cO(\vx) 
\right] \rangle_{alt}^{f_0} = \int \cD\phi_0~{\rm
   exp}[iI_{f_0}(\phi_0)]
\label{2-3}
\een
where
\ben
iI_{f_0}(\phi_0) = iS_{on}^\prime(\phi_0) +  \int
d^dx \left[\Cz(\epsilon) \frac{J(\vx)}{2\nu}
  \phi_0(\vx) -\Cz(\epsilon)^2 \frac{f_0}{2}
\left(\frac{\phi_0(\vx)}{2\nu}\right)^2 \right] 
\label{2-4}
\een
Using (\ref{1-10b}) and
performing the integral over $\phi_0$ this leads to the prediction
that the deformed CFT has a Green's function
\ben
G_{f}(k) = \frac{G_{alt}(k)}{1 + f_0 G_{alt}(k)}
\label{2-4a}
\een
This relation can be of course obtained directly from the large-N
field theory (\ref{double}) (see Eq. (\ref{green-f})). 
The holographic derivation of this formula is
a consistency check on the above dS/CFT prescription.

\section{\label{sec:hol-rg}Holographic RG}

We now adapt the holographic renormalization group procedure developed in
\cite{Heemskerk:2010hk, Faulkner:2010jy} to de Sitter space. 
we rewrite the right hand side
of (\ref{1-3})  by introducing a floating cutoff at $T=l$,
\ben
\Psi[\phi_0(\vx),\epsilon] = \int \cD\tphi (\vx) \Psi_{IR}[\tphi,l]
\Psi_{UV}[\tphi,\phi_0]
\label{1-6}
\een
where
\ben
 \Psi_{IR}[\tphi] = \Psi[\tphi(\vx),l]
\label{1-7}
\een 
and
\ben
\Psi_{UV}[\tphi,\phi_0] = \int^{\phi(\epsilon,\vx) = \phi_0(\vx)}_{\phi(l,\vx) = \tilde{\phi}(\vx)} \cD
\phi(T,\vx) {\rm{exp}}\left( i\int_{l}^\epsilon dT ~L \right)
\label{1-8}
\een
where $L$ is the Lagrangian. 

The idea is now to obtain an effective
action of the dual theory at a finite cutoff $l$ by extending the
dS/CFT relationship (\ref{1-14b}) for $\Psi_{IR} [\tphi,l]$,
\ben
\lan e^{-S_{eff}(l)}\rangle_{alt} = \int \cD \tphi (\vx) \int \cD J (\vx)~
\Psi_{UV}[\tphi,\phi_0]
{\rm{exp}} \left[ - \Cz(l)\kern-5pt \int d^dx \frac{J(\vx)}{2\nu}
  \tphi(\vx) \right] \nn \\  
\lan {\rm{exp}} \left[ \int d^dx J(\vx) \cO_- (\vx) \right] \ran_{alt}
\label{3-1}
\een
where $\Cz(l)$ is defined as in \eq{1-4}, with $\epsilon$ replaced by  $l$. 
This relates the parameters in $\Psi_{UV}$ to couplings in the
effective action. The expression for $\lan e^{ \int d^dx J(\vx) \cO_-
  (\vx) } \ran_{alt}$ in terms of bulk quantities in (\ref{1-10a}) and
(\ref{1-13}) are valid in the $\epsilon \rightarrow 0$ limit. When we
use these expressions for finite $l$, there is a freedom of choosing
counterterms \cite{Dong:2012afa, Balasubramanian:2012hb}. 
We will stick to the counterterm implied in
(\ref{1-10a}), and comment on the implications of this freedom later.

From the definition (\ref{1-8}), $\Psi_{UV}$ satisfies a Schrodinger
equation with the Hamiltonian derived from the Lagrangian, 
\ben 
iG_N
\frac{\partial}{\partial (-l)} \Psi_{UV} (\tphi,l) = H(l) \Psi_{UV}
(\tphi,l)
\label{3-2}
\een
which give flow equations for the parameters in $\Psi_{UV}$ and hence
couplings in the effective action. The negative sign in the left hand side of (\ref{3-2}) comes because time evolution corresponds to decreasing $l$, which appears as the lower limit of integration in (\ref{1-8}).

For the free scalar field we are considering the hamiltonian at some
time slice $T$ is given by 
\ben H(T) =\frac{1}{2} \int d^dx \left[ -
  G_N^2\left(\frac{-T}{\lds}\right)^{d-1} \frac{\delta^2}{\delta
    \phi^2} + \left( \frac{\lds}{-T} \right)^{d-1} (\nabla \phi)^2 +
  \left( \frac{\lds}{-T} \right)^{d+1}m^2 \phi^2 \right]
\label{3-3}
\een
In the semiclassical limit $G_N \ll \lds^{d-1}$ the Schrodinger equation 
reduces to a Hamilton-Jacobi equation. For a wavefunction 
\ben
\Psi_{UV} = {\rm exp} [iK]
\label{3-3a}
\een
the Hamilton-Jacobi equation is given by
\ben
\frac{1}{2}\left[ G_N^2 \left(\frac{-l}{\lds}\right)^{d-1}\left( \frac{\delta K}{\delta \phi} \right)^2 + \left( \frac{\lds}{-l} \right)^{d-1} (\nabla \phi)^2 + \left( \frac{\lds}{-l} \right)^{d+1}m^2 \phi^2 \right] + G_N \frac{\partial K}{\partial (-l)} = 0
\label{3-4}
\een
Consider now a general quadratic form for $K$
\ben
K =\frac{1}{G_N}\left( \frac{\lds}{-l} \right)^{d}\int d^d x \left[ - \frac{1}{2 \lds}g(l) \tphi^2 + h(l) \tphi +c (l)\right]
\label{3-5}
\een
Note that the parameters in (\ref{3-5}) depend on the cutoff $l$.
The flow equations for these parameters follow from
substituting (\ref{3-5}) in (\ref{3-4}). For consistency we really
need to replace these parameters by space-dependent parameters
(e.g. $g(x)$). However as shown in \cite{Heemskerk:2010hk} and
\cite{Elander:2011vh} the flow equations for the zero momentum
modes of these couplings decouple from the non-zero momentum
modes. With this understanding,
\bea
\beta_g & = & - (-il)\frac{\partial g}{\partial (-il)}  
=  -g^2 -dg - m^2\lds^2  \nn
\\
\beta_h &=& -(-il)\frac{\partial h}{\partial (-il)}  =  - h ( g + d )
\label{3-8}
\eea
As is clear from the discussion of \cite{Dong:2012afa} and
\cite{Balasubramanian:2012hb}, the freedom of choosing different
counterterms at finite $l$ modifies the last term in the first
equation of (\ref{3-8}). We have written the equations (\ref{3-8}) using $(-il)$ as a cutoff scale. This is a natural choice (as will be discussed further below).

The zeroes of $\beta_g$ are at $g_\pm = -\Delta_\pm$ and alternative
quantization means we have to expand the coupling as 
\ben
g = g_- + \delta g
\label{3-9}
\een
The beta function for $\delta g$ is given by
\ben
\beta_{\delta g} = -(-il)\frac{\partial \delta g}{\partial (-il)}= -2\nu
(\delta g) - (\delta g)^2
\label{3-9a}
\een

To relate this flow equations to beta functions of the dual field theory we need to establish a relationship between $g, f$ and the couplings of the field theory. This may be done by substituting (\ref{3-5}) in (\ref{3-1}) and performing the integrals over $J(\vx)$ and $\tphi(\vx)$ by saddle point method.  This leads to a field theory effective action
\ben
S_{eff} = \frac{f}{2}\int d^d x~ \cO_-^2 + j\int d^d x~ \cO_- + c
\label{3-6}
\een
where
\bea
j &=& -2\nu \sqrt{\frac{L_{dS}^{d+1} \gamma}{G_N}} (i)^{d+1} (-il)^{-\Delta_+} h(l) \\
f &=& (i)^{d+1}(-il)^{-2\nu} (2\nu)^2 g~\gamma = - (2\nu)^2 C(\nu)
\frac{1}{b_{dS}}~(-il)^{-2\nu}~g
\label{3-7}
\eea
and $c$ is a constant independent of the operator $\cO$. In the above we have used the expression for $b_{dS}$ in (\ref{1-14a}).

The fixed point values of the parameter $g$ simply corresponds to the minimal counterterm in the bulk action . The field theory couplings, which are defined as departures from a CFT have to be related to the departure from the fixed point.

The couplings $f,j$ and hence $\delta f$ and $\delta j$  have the appropriate dimensions $2\nu$ and $\Delta_+$ respectively, as is clear from the powers of $l$ which appear in (\ref{3-7}).
The beta functions of the field theory are, however, those of {\em dimensionless} couplings. In the field theory this is done by multiplying by an appropriate power of the cutoff or renormalization scale, as in (\ref{dim-less}).
In the holographic setup this requires specifying a relationship between the cutoff in the bulk with a UV cutoff on the boundary. As is quite clear from all the formulae above, it is natural to identify $(-il)$ as the
renormalization scale $\mu$ of the field theory. Let us identify the field
theory renormalization scale $\mu$ to be $a^{1/(2\nu)}$ times
the holographic cut-off scale $1/(-il)$, 
for some positive constant $a$. With this choice,
we have  the following identification of the dimensionless coupling of the field theory $\lambda$ with the departure from the fixed point,
\ben
\lambda \, a \left((-il)^{-1}\right)^{2\nu} \equiv  \delta f 
= - (2\nu)^2 C(\nu) \frac{1}{b_{dS}}~
(-i l)^{-2\nu}~\delta g
\een 
where we have used \eq{dim-less}. Making the convenient choice
$a = (2\nu)^3 C(\nu)$ (which gives a specific choice of the
field theory renormalization scale), we get
\ben 
\delta g =- 2\nu b_{dS} \,\lambda 
\label{3-10}
\een
Substituting this in (\ref{3-9a}) finally leads to a beta function 
for $\lambda$
\ben
\beta_\lambda = - 2\nu \lambda + 2\nu\, b_{dS} \lambda^2
= - 2\nu \lambda + 2\nu\, \frac{i^{1-d}}{\gamma} C(\nu) \lambda^2
\label{3-10b}
\een
This is the same as the general field theory answer, (\ref{beta}). 

As we have remarked above and will discuss in detail in Section
\ref{sec:complex}, the requirement that there are no relative phases
between various $n$-point functions of the dual field theory
\footnote{\label{ftnt:all-real}It is clear from our discussion in Section \ref{sec:complex} that under no circumstance can the $n$-point functions 
be all real.}  implies
that $b_{dS} \sim i^{d-1}$. This implies, in turn, that for even $d$
we have purely imaginary $b_{dS}$ and hence a complex beta function.

\subsection{Results in $AdS$}

For comparison let us recall the results of the above analysis in euclidean $AdS$. In this case the range of the radial coordinate is $0 \leq z \leq \infty$. The radial evolution equation satisfied by $\Psi_{UV.AdS}$ is
 \ben 
G_N
\frac{\partial}{\partial (l)} \Psi_{UV,AdS} (\tphi,l) =- H_{AdS}(l) \Psi_{UV,AdS}
(\tphi,l)
\label{3-2b}
\een
where 
\ben H(l) =\frac{1}{2} \int d^dx \left[ -
  G_N^2\left(\frac{l}{L_{AdS}}\right)^{d-1} \frac{\delta^2}{\delta
    \phi^2} + \left( \frac{L_{AdS}}{l} \right)^{d-1} (\nabla \phi)^2 +
  \left( \frac{L_{AdS}}{l} \right)^{d+1}m^2 \phi^2 \right]
\label{3-3b}
\een
With the form 
\ben
\Psi_{UV,AdS}= {\rm exp} \left[ \frac{1}{G_N}\left( \frac{L_{AdS}}{l} \right)^{d}\int d^d x \left[ - \frac{1}{2 L_{AdS}}g^\prime(l) \tphi^2 + h^\prime(l) \tphi +c^\prime(l)\right] \right]
\een
which leads to the flow equation
\ben
l\frac{\partial g^\prime}{\partial l}  
=  -(g^\prime)^2 -dg^\prime + m^2 L_{AdS}^2
\een
The expressions for the fixed points are changed appropriately, but the flow equation for the departure from the fixed point $\delta g^\prime$
is, instead of (\ref{3-9a})
\ben
\beta_{\delta g^\prime} = -l\frac{\partial \delta g^\prime}{\partial l}= -2\nu
(\delta g^\prime) + (\delta g^\prime)^2
\label{3-9b}
\een
Finally the relationship between the field theory dimensionless coupling and $\delta g^\prime$ is
\ben
\lambda =  \gamma (2\nu)^2~\delta g' = (2\nu) 2^{2\nu} \frac{\Gamma(1+\nu)}{\Gamma(1-\nu)}\frac{1}{b_{AdS}} \delta g^\prime
\label{3-10d}
\een
which leads once again to a beta function of the expected form (\ref{beta})

\section{Beta function of Triple and Higher trace couplings}

In this section we will discuss a generalization of the above methods
to derive the holographic beta-function of triple and higher trace
couplings (in perturbation theory). We will be brief, emphasizing mainly
the new features. 

For concreteness, we will focus on triple trace couplings, of the
form $\cO_-^3$; however, the generalization
to higher trace operators is straightforward.  
Triple trace operators are induced in a holographic RG, as we will see,
when the dual scalar field theory has a cubic coupling
\ben
\Delta S_\epsilon = \frac{1}{G_N}\int_{-\infty}^\epsilon dT \int d^{d}x \rat^{d+1} 
\left[ - \frac r 3 \phi^3 \right]
\label{1-2a}
\een
in additional to the quadratic action \eq{1-2}. The Hamilton-Jacobi
equation \eq{3-4} gets modified by the addition of a cubic term
\[
\left( \frac{\lds}{-l} \right)^{d+1} \frac{2 r}3 \phi^3
\]
to the term inside the square bracket. It is easy to see that
a quadratic ansatz for the kernel $K$ such as \eq{3-5} will not satisfy such a Hamilton-Jacobi
equation. Let us, therefore, take $K$ to be cubic, {\em viz.} of the form
\ben
K = \rho^2 L \ep^{-d} \int d^d x \bigg( -\frac{g}{2L} \tphi^2 + h\tphi + c L +\frac{A}{L^2}\frac{\tphi^3}{3}\bigg) 
\label{cubic-K}
\een
By repeating the steps leading to \eq{3-8}, and equating
the coefficients of $\tphi, \tphi^2$ and $\tphi^3$ in the Hamilton-Jacobi
equation,\footnote{Our approach here is perturbative; the Hamilton-Jacobi
analysis generates $\tphi^4$ terms. We imagine them to
be taken care of by higher couplings, and focus here on couplings
up to cubic order. It is straightforward, although cumbersome, to write
more general beta-functions involving arbitrary Wilsonian
couplings.}  we now get the
following cut-off dependence of the couplings in \eq{cubic-K}
\begin{align}
\b_g &= - g^2 - d~g - {\bar m}^2  - 2 h A, ~~~ {\bar m}=m L_{dS},
\nn\\
\b_A &=  (-3g - d) A + 3\bar r, ~~~ {\bar r}=  r L_{dS}^3,
\nn\\
\b_h &= (-g-d) h
\label{beta-cubic}
\end{align}
Note that this generalizes \eq{3-8}, and reduces to it
for $A=0$. 
It is easy to find the following UV fixed point (near
which $\b_g$ is negative):
\begin{align}
h_c=0, ~ g_c= -\Delta_- ,~   A_c= 3 \bar r/(d - 3\Delta_-)
\label{fixed}
\end{align}
The linearized beta-functions for the deformations $\delta h, \delta g$
and $\delta A$ (measured from this fixed point) are
\begin{align}
\b_{\delta g}= -2\nu \delta g, ~~~~
\b_{\delta A}=  (3 \Delta_-  - d) \delta A ,~~~~
\b_{\delta h}=  (\Delta_-- d)\delta h 
\label{beta-cubic-linear}
\end{align}  
How does one read off the field theory beta-functions from these?
We can, once again, use \eq{3-1}, and show that it leads
to a field theory with the following effective action
\ben
S_{eff} = \int d^d x \left(\frac{f}{2} \cO_-^2 + j \cO_- + 
\frac{B}{3} \cO_-^3 + c  \right)
\label{cubic-CFT}
\een
where
\bea
j &=& -(i)^{d+1} (-il)^{-\Delta_+} h(l) 2\nu \sqrt{\frac{L_{dS}^{d+1} \gamma}{G_N}}  
\nonumber\\
f &=& (i)^{d+1}(-il)^{-2\nu} (2\nu)^2 g(l) ~\gamma 
\nonumber\\
B &=& - (i)^{d+1} (-il)^{3\Delta_--d} (2\nu)^3 A(l) \gamma \sqrt{\frac{\gamma~G_N }{L_{dS}^{d+1}}}
\label{cubic-identify}
\eea
which generalizes the equation \eq{3-7} encountered for double trace
couplings. The beta-function for the field theory couplings $f,j, B$ can easily
be read off from the above identifications \eq{cubic-identify} with the bulk couplings $g,h, A$
and their beta-functions \eq{beta-cubic} or \eq{beta-cubic-linear}. The beta function for the dimensionless cubic trace coupling ($\delta\bar{B}$),
which measures the deviation from the fixed point, turns out to be,
\ben
\beta_{\delta\bar{B}} = 
-3 \Delta_- \delta\bar{B} + 3 \frac{i^{1-d}}{(2\nu)^2\gamma}\delta \bar{f} 
\delta\bar{B} =  -3 \Delta_- \delta\bar{B} + 3~ b_{dS}~ 
\delta\bar{f}\, \delta\bar{B} ~ \frac{\Gamma(1-\nu)}{2^{2\nu}(2\nu)\Gamma(1+\nu)}
\een
where, $\delta\bar{f}$ is the deviation of
the dimensionless double trace coupling
from the fixed point. One can easily check that the field theory beta-functions have the correct form. E.g., $\beta_{\delta\bar B}$ 
includes a term $\propto \delta\bar f \delta \bar B$;
to see this from a field theory reasoning, 
one needs to simply note that the three-point function
$\lan \cO_- (x)\cO_- (y)\cO_- (z) \ran$ has a perturbative expansion of the
schematic form $\delta \bar B \int d^d w  G_0(x-w) G_0(y-w) G_0(z-w) 
+  \delta \bar f
\delta \bar B \int d^d w~ d^d w'  G_0(x-w) G_0(y-w) G_0(z-w')G_0(w-w')$
(where we have shown only the first two terms). Using
large $N$ methods, one can organize such perturbation expansions
\cite{Witten:2001ua, Mueck:2002gm, Vecchi:2010dd}. 

Significantly, the beta function for $A$ does not have an $A^2$ term
(in field theory terms, $\beta_{\delta \bar B}$ does not have a 
${\delta \bar B}^2$ term), 
and is in fact the same as in $AdS$. 
For the special case where $d=3, \Delta_- =1$
(which implies ${\bar m}^2=2, \Delta_+ =2, \nu =1/2$)  the linearized
beta-function indicates correctly the fact that the cubic coupling
is  marginal \footnote{The fixed point value of $A$ is infinite
for these values, as can be seen from \eq{fixed}. However, as
remarked earlier, the fixed point value of holographic couplings
is non-universal as they are affected by the choice of 
holographic counterterms. The linearized beta-functions \eq{beta-cubic-linear}
are free of such non-universalities. 
}. This is consistent with the known field theory result for vector models that a $[(\vec{\phi})^2]^3$ coupling acquires a nontrivial beta function only due to $1/N$ corrections. Our holographic result shows that this is a general result in large-N field theories.

\section{\label{sec:complex}Complex Phases}

Here we focus on the  structure of complex
phases of the $n$-point correlation functions of the field theory. As seen in \cite{strom1} even with interactions present in the bulk, the overall factor in $iI_{on-shell}$ is $i^{d-1}$. This implies the following schematic relations for leading order contributions to the first few $n$-point correlation functions,
\begin{align}
&  i^{d-1} = \Cz^2 \lan O O \ran =  \gamma^{-1}  b_{dS}
\nn\\
&  r_3 i^{d-1}=  \Cz^3  \lan O O O \ran
\nn\\
 &  r_4 i^{d-1}=  \Cz^4  \lan O O O O\ran
\label{a-1}
\end{align}
In these equations, we display only those quantities which possibly
contain complex phases. The quantity $\Cz$ is defined in \eq{1-4};
since $-i\epsilon$ has been identified with a real cut-off of the
field theory, i.e. $-i\epsilon \propto 1/\Lambda_{UV}$, $\Cz$
is essentially equal to $1/\sqrt \gamma$ so far as keeping track
of  complex phases is concerned. Similarly, we have written $\lan 
O O \ran \propto b_{dS}$.
The left hand sides of the above set of equations are obtained from the bulk; e.g. the LHS of the top equation displays the
complex phase of \eq{1-10b}. The couplings $r_3$, $r_4$ 
represent cubic, quartic, etc. couplings of the scalar Lagrangian (e.g.
$r_3$ is the same as $r$ in \eq{1-2a}). In keeping with unitarity of
the bulk field theory, we will assume that these coefficients are all
real. The right hand sides of 
equations \eq{a-1} are obtained by the GKPW prescription, i.e.
by expanding $\Psi[\phi_0(x), \epsilon]$ in
\eq{1-4} in powers of $\phi_0(x)$.  Now, if we require that there is no relative phase between the correlation functions, i.e, the phase of $\lan O_1 \cdots O_n \ran$ = the phase of $\lan O_1 \cdots O_{n+1} \ran$, then we must have $\Cz$ real. Recalling that $\Cz(\epsilon) \propto 1/\sqrt \gamma$ (where
the proportionality constant is positive),
the reality of $\Cz$ implies that $\gamma$ is real. 
Thus, so far as keeping track of complex
phases is concerned, it can be taken to be $1$. It then follows immediately that $b_{dS}$ is complex for even $d$ leading to complex beta functions. \\ 
Alternatively if we want to require the beta function to be always real, i.e, $b_{dS}$ to be real, then we must choose the phase of $\Cz$ 
to be $i^{(d-1)/2}$. However since the phases of the left hand sides of \eq{a-1} 
are all equal,  this will now imply the following relative complex phase,
\ben
\lan O_1 \cdots O_{n+1} \ran = i^{(1-d)/2}\lan O_1 \cdots O_{n} \ran
\label{phase}
\een
In particular, since in this choice $\lan O O \ran$ is real, we get that the phase of $\lan O_1 \cdots O_{n} \ran$ is $i^{(n-2)(1-d)/2}$. This clearly shows that we cannot have both the beta function as real and the absence of $n$-dependent phases.

\section{Note added}

While this paper was in the final stages of its preparation, \cite{anninos3} appeared on the archive, which contains a discussion of the effect of double trace deformations of the free $Sp(N)$ theory in $2+1$ dimensions.

\section*{Acknowledgements} We would like to thank Ganpathy Murthy and Sandip Trivedi for
discussions and T. Hartman for a useful correspondence. S.R.D. thanks Tata Institute of Fundamental Research for hospitality during the initial stages of this work. The work of D.D. and S.R.D. was partially supported by a
grant from National Science Foundation NSF-PHY-1214341.


\begin{thebibliography}{99}

\baselineskip=10pt

\bibitem{Witten:2001kn} 
  E.~Witten,
  ``Quantum gravity in de Sitter space,''
  hep-th/0106109.

\bibitem{Strominger:2001pn} 
  A.~Strominger,
  ``The dS / CFT correspondence,''
  JHEP {\bf 0110}, 034 (2001)
  [hep-th/0106113].


\bibitem{Maldacena:2002vr} 
  J.~M.~Maldacena,
  ``Non-Gaussian features of primordial fluctuations in single field inflationary models,''
  JHEP {\bf 0305}, 013 (2003)
  [astro-ph/0210603].

\bibitem{Harlow:2011ke} 
  D.~Harlow and D.~Stanford,
  ``Operator Dictionaries and Wave Functions in AdS/CFT and dS/CFT,''
  arXiv:1104.2621 [hep-th].


\bibitem{Maldacena}
  J.~M.~Maldacena,
 ``The large N limit of superconformal field theories and supergravity,''
  Adv.\ Theor.\ Math.\ Phys.\  {\bf 2} (1998) 231
  [Int.\ J.\ Theor.\ Phys.\  {\bf 38} (1999) 1113]
  [arXiv:hep-th/9711200];


\bibitem{GKP}
S.~S.~Gubser, I.~R.~Klebanov and A.~M.~Polyakov,
  ``Gauge theory correlators from non-critical string theory,''
  Phys.\ Lett.\ B {\bf 428}, 105 (1998)
  [arXiv:hep-th/9802109].

\bibitem{W}
E.~Witten,
 ``Anti-de Sitter space and holography,''
  Adv.\ Theor.\ Math.\ Phys.\  {\bf 2}, 253 (1998)
  [arXiv:hep-th/9802150].


\bibitem{AdSR}
O.~Aharony, S.~S.~Gubser, J.~M.~Maldacena, H.~Ooguri and Y.~Oz,
``Large N field theories, string theory and gravity,''
  Phys.\ Rept.\  {\bf 323}, 183 (2000)
  [arXiv:hep-th/9905111].



\bibitem{strom1} 
  D.~Anninos, T.~Hartman, A.~Strominger and ,
  ``Higher Spin Realization of the dS/CFT Correspondence,''
  arXiv:1108.5735 [hep-th].

\bibitem{ElShowk:2011ag} 
  S.~El-Showk and K.~Papadodimas,
  ``Emergent Spacetime and Holographic CFTs,''
  JHEP {\bf 1210}, 106 (2012)
  [arXiv:1101.4163 [hep-th]].

\bibitem{Ng} 
  G.~S.~Ng, A.~Strominger and ,
  ``State/Operator Correspondence in Higher-Spin dS/CFT,''
  arXiv:1204.1057 [hep-th].

\bibitem{Anninos} 
  D.~Anninos,
  ``De Sitter Musings,''
  Int.\ J.\ Mod.\ Phys.\ A {\bf 27}, 1230013 (2012)
  [arXiv:1205.3855 [hep-th]].

\bibitem{ddjy} 
  D.~Das, S.~R.~Das, A.~Jevicki, Q.~Ye and ,
  ``Bi-local Construction of Sp(2N)/dS Higher Spin Correspondence,''
  JHEP {\bf 1301}, 107 (2013)
  [arXiv:1205.5776 [hep-th]].

\bibitem{Anninos2} 
  D.~Anninos, F.~Denef, D.~Harlow and ,
  ``The Wave Function of Vasiliev's Universe - A Few Slices Thereof,''
  arXiv:1207.5517 [hep-th].

\bibitem{Witten:2001ua} 
  E.~Witten,
  ``Multitrace operators, boundary conditions, and AdS / CFT correspondence,''
  hep-th/0112258.

\bibitem{Klebanov:1999tb} 
  I.~R.~Klebanov and E.~Witten,
  ``AdS / CFT correspondence and symmetry breaking,''
  Nucl.\ Phys.\ B {\bf 556}, 89 (1999)
  [hep-th/9905104].


\bibitem{Mueck:2002gm} 
  W.~Mueck,
  ``An Improved correspondence formula for AdS / CFT with multitrace operators,''
  Phys.\ Lett.\ B {\bf 531}, 301 (2002)
  [hep-th/0201100].


\bibitem{Gubser:2002vv} 
  S.~S.~Gubser and I.~R.~Klebanov,
  ``A Universal result on central charges in the presence of double trace deformations,''
  Nucl.\ Phys.\ B {\bf 656}, 23 (2003)
  [hep-th/0212138].

\bibitem{Vecchi:2010dd} 
  L.~Vecchi,
  ``The Conformal Window of deformed CFT's in the planar limit,''
  Phys.\ Rev.\ D {\bf 82}, 045013 (2010)
  [arXiv:1004.2063 [hep-th]].
  L.~Vecchi,
  ``Multitrace deformations, Gamow states, and Stability of AdS/CFT,''
  JHEP {\bf 1104}, 056 (2011)
  [arXiv:1005.4921 [hep-th]].


\bibitem{Heemskerk:2010hk} 
  I.~Heemskerk and J.~Polchinski,
  ``Holographic and Wilsonian Renormalization Groups,''
  JHEP {\bf 1106}, 031 (2011)
  [arXiv:1010.1264 [hep-th]].

\bibitem{Faulkner:2010jy} 
  T.~Faulkner, H.~Liu and M.~Rangamani,
  ``Integrating out geometry: Holographic Wilsonian RG and the membrane paradigm,''
  JHEP {\bf 1108}, 051 (2011)
  [arXiv:1010.4036 [hep-th]].

\bibitem{Akhmedov:1998vf}
  E.~T.~Akhmedov,
  ``A Remark on the AdS / CFT correspondence and the renormalization group flow,''
  Phys.\ Lett.\  {\bf B442 } (1998)  152-158.
  [hep-th/9806217].
  
\bibitem{Alvarez:1998wr}
  E.~Alvarez and C.~Gomez,
  ``Geometric holography, the renormalization group and the c theorem,''
  Nucl.\ Phys.\  B {\bf 541}, 441 (1999)
  [arXiv:hep-th/9807226].

\bibitem{Balasubramanian:1999jd}
  V.~Balasubramanian and P.~Kraus,
  ``Space-time and the holographic renormalization group,''
  Phys.\ Rev.\ Lett.\  {\bf 83}, 3605 (1999)
  [arXiv:hep-th/9903190].

\bibitem{Freedman:1999gp}
  D.~Z.~Freedman, S.~S.~Gubser, K.~Pilch and N.~P.~Warner,
   ``Renormalization group flows from holography supersymmetry and a c
  theorem,''
  Adv.\ Theor.\ Math.\ Phys.\  {\bf 3}, 363 (1999)
  [arXiv:hep-th/9904017].

\bibitem{deBoer:1999xf}
  J.~de Boer, E.~P.~Verlinde and H.~L.~Verlinde,
  ``On the holographic renormalization group,''
  JHEP {\bf 0008}, 003 (2000)
  [arXiv:hep-th/9912012].

\bibitem{Bianchi:2001kw}
  M.~Bianchi, D.~Z.~Freedman, K.~Skenderis,
  ``Holographic renormalization,''
  Nucl.\ Phys.\  {\bf B631 } (2002)  159-194.
  [hep-th/0112119].

\bibitem{Akhmedov:2002gq}
  E.~T.~Akhmedov,
  ``Notes on multitrace operators and holographic renormalization group,''
  [hep-th/0202055].


\bibitem{Skenderis:2002wp}
  K.~Skenderis,
  ``Lecture notes on holographic renormalization,''
  Class.\ Quant.\ Grav.\  {\bf 19}, 5849 (2002)
  [arXiv:hep-th/0209067].


\bibitem{Elander:2011vh} 
  D.~Elander, H.~Isono and G.~Mandal,
  ``Holographic Wilsonian flows and emergent fermions in extremal charged black holes,''
  JHEP {\bf 1111}, 155 (2011)
  [arXiv:1109.3366 [hep-th]].

\bibitem{Balasubramanian:2012hb} 
  V.~Balasubramanian, M.~Guica and A.~Lawrence,
  ``Holographic Interpretations of the Renormalization Group,''
  JHEP {\bf 1301}, 115 (2013)
  [JHEP {\bf 1301}, 115 (2013)]
  [arXiv:1211.1729 [hep-th]].

\bibitem{Dong:2012afa} 
  X.~Dong, B.~Horn, E.~Silverstein and G.~Torroba,
  ``Moduli Stabilization and the Holographic RG for AdS and dS,''
  arXiv:1209.5392 [hep-th].

\bibitem{LeClair:2007iy} 
  A.~LeClair and M.~Neubert,
  ``Semi-Lorentz invariance, unitarity, and critical exponents of symplectic fermion models,''
  JHEP {\bf 0710}, 027 (2007)
  [arXiv:0705.4657 [hep-th]].

\bibitem{anninos3} 
  D.~Anninos, F.~Denef, G.~Konstantinidis and E.~Shaghoulian,
  ``Higher Spin de Sitter Holography from Functional Determinants,''
  arXiv:1305.6321 [hep-th].

\end{thebibliography}
\end{document}